\documentclass[10pt, conf, twocolumn]{IEEEtran}
\usepackage{cleveref}
\usepackage{color}
\usepackage{times}
\usepackage{epsfig}
\usepackage{graphicx}
\usepackage{epstopdf}
\usepackage{algorithm}
\usepackage{algorithmic}
\usepackage{amsmath}
\usepackage{amssymb}
\usepackage{bbm}
\usepackage{amsxtra}
\usepackage{multirow}
\usepackage{mathtools}
\usepackage{caption}
\usepackage{cleveref}
\usepackage{url}
\usepackage{float}
\usepackage[font=small, singlelinecheck=false]{caption}
\usepackage{caption}
\usepackage[caption = false]{subfig}
\usepackage{cite}
\usepackage{amsthm}

\newtheorem*{proof*}{Proof}

\begin{document}

\title{Cognitive Connectivity Resilience in Multi-layer Remotely Deployed Mobile Internet of Things}

\author{ \IEEEauthorblockN{\large Muhammad Junaid Farooq and Quanyan Zhu} \\ \IEEEauthorblockA{Department of Electrical \& Computer Engineering, Tandon School of Engineering, \\New York University, Brooklyn, NY, USA,} Emails: \{mjf514, qz494\}@nyu.edu. \vspace{-0.2in}
%{\thanks {\vspace{-0.2cm}\hrule \vspace{0.2cm} \indent This work was made possible by.}}
}

\maketitle

\begin{abstract}
Enabling the Internet of things in remote areas without traditional communication infrastructure requires a multi-layer network architecture. The devices in the overlay network are required to provide coverage to the underlay devices as well as to remain connected to other overlay devices. The coordination, planning, and design of such two-layer heterogeneous networks is an important problem to address.  Moreover, the mobility of the nodes and their vulnerability to adversaries pose new challenges to the connectivity. For instance, the connectivity of devices can be affected by changes in the network, e.g., the mobility of the underlay devices or the unavailability of overlay devices due to failure or adversarial attacks. To this end, this work proposes a feedback based adaptive, self-configurable, and resilient framework for the overlay network that cognitively adapts to the changes in the network to provide reliable connectivity between spatially dispersed smart devices. Our results show that if sufficient overlay devices are available, the framework leads to a connected configuration that ensures a high coverage of the mobile underlay network. Moreover, the framework can actively reconfigure itself in the event of varying levels of device failure.
%We develop a distributed and dynamic framework for the overlay network that aims to maximize the coverage of the underlay network while remaining interconnected at the same time.

%Enabling the Internet of things in remote areas without communication infrastructure requires a multi-layer communication network. The overlay network is required to provide coverage to the underlay devices as well as remain connected to other overlay devices which makes it a challenging problem from a network planning perspective. Moreover, the connectivity of the devices can be affected by changes in the network, e.g., the mobility of the underlay devices or the unavailability of overlay devices due to failure or adversarial attacks. This necessitates an adaptive, self-configurable, and resilient framework for the overlay network overlay network that cognitively adapts to the changes in the network to provide reliable connectivity between spatially dispersed smart devices. We develop a distributed and dynamic framework for the overlay network hat aims to maximize the coverage of the underlay network while remaining interconnected at the same time. Our results show that if sufficient overlay devices are available, the framework leads to a connected lattice configuration that maximizes the ensures a high coverage of the mobile underlay network. Moreover, the framework is able to actively reconfigure itself in the event of varying levels of random device failure.
\end{abstract}

\IEEEpeerreviewmaketitle


\vspace{-0.0in}
\section{Introduction}
Connectivity is vital in enabling the emerging paradigm of the Internet of things (IoT)~\cite{iot_ref}. The fundamental objective of IoT is to inter-connect smart objects together so that they can exchange data and leverage the capabilities of each other for achieving individual or network goals such as high efficiency, accuracy and economic benefits. Traditionally, IoT devices are connected to an access point which, in-turn, is connected to a wired or wireless backhaul network. The backhaul network enables connectivity and accessibility between things that are geographically separated. However, backhaul networks may not always be available such as in remote areas~\cite{iort}, disaster struck areas~\cite{disaster}, and battlefields~\cite{iobt_junaid}. Unmanned aerial vehicles (UAVs) and mobile ground stations have been the most viable candidates for providing connectivity in such situations. In fact, there is a growing interest towards the use of drones and UAVs as mobile aerial base stations (BSs) to assist existing cellular LTE networks~\cite{drone_placement}, public safety networks~\cite{drone_public_safety}, and intelligent transportation systems~\cite{uav_its}.

\begin{figure}
  \centering
  \includegraphics[width=2.5in]{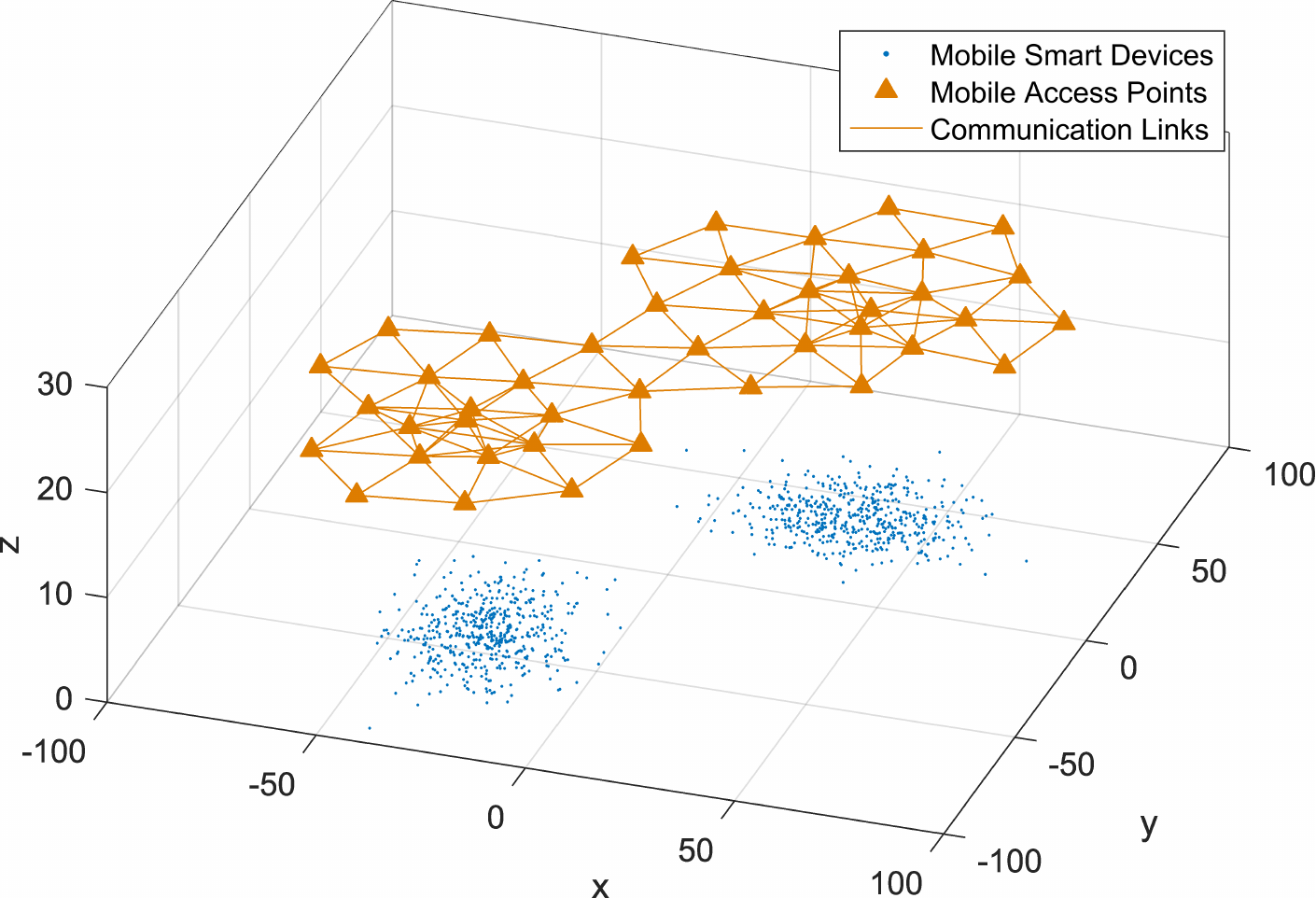}\\
  \caption{Example of spatially clustered mobile smart devices interconnected by an overlay network of mobile access points.\vspace{-0.0in}}\label{remote_iot_fig}
\end{figure}

Due to the absence of traditional communication infrastructure and backhaul networks, the remotely deployed IoT requires a multi-layer architecture comprising of an overlay network of mobile access points (MAPs) to interconnect the spatially dispersed mobile smart devices (MSDs). The MAPs exploit device-to-device (D2D) communications~\cite{d2d_iot} for connecting with other MAPs while the MSDs connect to one of the available MAPs for communication. The problem in such settings is to efficiently deploy the overlay network that provides coverage to all the MSDs as well as maintaining connectivity between the MAPs. Since the MSDs can be located in spatial clusters that are arbitrarily separated, the MAPs should be deployed in a way that they remain connected, i.e., each MAP is reachable from other MAPs using D2D communications. This requirement makes it a challenging network planning and design problem. Fig.~\ref{remote_iot_fig} illustrates one such scenario where the MAPs are appropriately deployed enabling a local inter-network of MSDs without any traditional communication infrastructure. It can be easily connected to the internet to achieve pervasive connectivity and control over the MSDs.

Several efforts have been made in literature towards efficient deployment of aerial BSs to serve ground users such as \cite{ot_uav_deployment,drone_placement}.
The power-efficient deployment of UAVs as aerial BSs is studied in~\cite{ot_uav_deployment} with the objective to collect data from remotely deployed sensors and not to inter-connect BSs in the air. In~\cite{drone_placement}, the authors have proposed a backhaul-aware deployment that is applicable to settings with traditional communication infrastructure.
%In~\cite{circle_packing}, the authors propose a circle packing solution for providing energy efficient coverage in an area. While the circle packing approach does enforce a regular arrangement of the UAVs, the applicability of the solution is limited as it is not scalable and also not well suited for an arbitrary spatial distribution of users.
Most existing works dealing with UAV placement formulate the BS deployment problem as a \emph{facility location} problem. However, the solution to the facility location problem is not sufficient to ensure the inter-connectivity of the facilities. In our case, the MAPs are wireless devices which have limited communication range and have to be located in sufficient proximity to communicate reliably. Our  goal is to place the MAPs in a connected configuration to enable inter-connectivity between the underlying MSDs using D2D links, which is unique to the wireless network setting. This problem is significantly more complex than the multi-facility location problem, also referred to as the \emph{p-median} problem~\cite{p_median}, that is known to be NP-hard. Hence, a globally optimal solution to this problem is not easy to obtain. Moreover, a centralized solution is also less attractive due to the practical limitations in the scenario considered in this paper since the two-layer network cannot be coordinated by a central planner.

In addition to effective initial deployment of MAPs, there is a need for an autonomic, self-organizing, and self-healing overlay network that can continuously adapt and reconfigure according to the constantly changing network conditions~\cite{chen_paper}. The MSDs can be highly mobile such as smart handheld devices, wireless sensors, and wearable devices whose mobility can be either individual or collective based on the objective such as a rescue operation or a battlefield mission. Furthermore, the network is also vulnerable to failures and cyber-physical adversarial attacks. Therefore, a distributed and dynamic approach to providing resilient connectivity is essential to cope with the growing scale of the networks towards a massive IoT~\cite{massive_iot}. To this end, we propose a feedback based distributed cognitive framework that maintains connectivity of the network and is resilient to the mobility of MSDs and/or failures of the MAPs. The continuous feedback enables the framework to actively react to network changes and appropriately reconfigure the network in response to a failure event that has resulted in loss of connectivity. Simulation results demonstrate that if sufficient MAPs are available, they can be arranged into a desired configuration from arbitrary initial positions and the configuration continuously adapts according to the movement of the MSDs as well as recovers connectivity under varying levels of a random MAP failure event.

The rest of the paper is organized as follows: Section~\ref{Sec:Sys_model} provides the system model describing the connectivity between the overlay and underlay networks. Section~\ref{Sec:Methodology} presents the proposed feedback based cognitive connectivity framework while Section~\ref{Sec:Perf_evaulation} defines the metrics used for performance evaluation. In Section~\ref{Sec:Results}, we provide results on the convergence of the framework and the resilience to mobility and random failures. Finally, Section~\ref{Sec:Conclusion} concludes the paper.

%One related work in literature is the recent development of multi-agent control methods to provide distributed coverage~\cite{coverage_control}. However, the direct application of Voronoi partitioning methods in~\cite{coverage_control} is not viable as the centroids of Voronoi cells may be arbitrarily separated resulting in a disconnected configuration that is not desirable in the D2D network. Another related work is the distributed formation control of autonomous agents that has been developed in the context of robot swarming~\cite{flocking}. One important distinct challenge of our problem is to take into account the spatial distribution of the underlying MSDs as well as the coverage capacity constraints of the overlay MAPs.

%Distributed coverage of underlying users by autonomous agents has been well studied in literature~\cite{coverage_control}. However, it is also based on Voronoi partitioning of the space according to the user density and does not take into account the connectivity of the Voronoi centers. The distributed formation control of autonomous agents in a lattice configuration has been developed in the context of \emph{flocking} of robots~\cite{flocking}. However, it does not take into account the spatial distribution of the underlying network as well as the coverage capacity constraints of the robots.

%D2D backhaul on the fly
\vspace{-0.0in}
\section{System Model}\label{Sec:Sys_model}
We consider a finite set of MSDs arbitrarily placed in $\mathbb{R}^2$ denoted by $\mathcal{M} = \{1, \ldots, M\}$ and a finite set of MAPs denoted by $\mathcal{L} = \{1, \ldots, L\}$, placed in $\mathbb{R}^3$ at a constant elevation\footnote{We assume MAPs have a constant elevation from the ground for simplicity, however, the methodology and results can be readily generalized to varying elevations.} of $h$, for providing connectivity to the MSDs. The MAPs have a maximum communication range of $r \in \mathbb{R}^+$, i.e., any two MAPs can communicate only if the Euclidean distance between them is less than $r$. The Cartesian coordinates of the MSDs at time $t$ are denoted by $\mathbf{y}(t) = [y_1(t), y_2(t), \ldots, y_M(t)]^T$, where $y_i(t) \in \mathbb{R}^2, \forall i \in \mathcal{M}, t \geq 0$. Similarly, the Cartesian coordinates of the MAPs at time $t$ are denoted by $\mathbf{q}(t) = [q_1(t), q_2(t), \ldots, q_{L}(t)]^T$, where $q_i \in \mathbb{R}^3$, $\forall i \in \mathcal{L}, t \geq 0$. For brevity of notation, we drop the time index henceforth and assume that the time dependence is implicitly implied. The communication neighbours of each MAP is represented by the set $\mathcal{N}_i = \{ j \in \mathcal{L}, j \neq i : \| q_i - q_j \| \leq r \}, \forall i \in \mathcal{L}$. The quality or strength of the communication links between the MAPs is modeled using a distance dependent decaying function $\alpha_{\{z_1,z_0\}}(z) \in [0,1]$ with finite cut-offs, expressed as follows:
\begin{align} \label{beta_func}
\alpha_{\{z_1, z_0\}}(z) =
\begin{cases}
1, & \text{if } 0 \leq z < z_1, \\
0.5 \left( 1 + \cos(\pi \frac{z - z_1}{z_0 - z_1}) \right), & \text{if } z_1 \leq z < z_0, \\
0, & \text{if } z > z_0,
\end{cases}
\end{align}
where $z_0$ and $z_1$ are constants that define the cut-off values corresponding to $0$ and $1$ respectively. Moreover, in order to make the norm measure of a vector differentiable at the origin, a new mapping of the L-2 norm is defined following~\cite{flocking}, referred to as the $\sigma-$norm\footnote{Note that this is not a norm but a mapping from a vector space to a scalar.}:
\begin{align}
\|\text{x}\|_{\sigma} = \frac{1}{\epsilon}\left( \sqrt{1 + \epsilon \| \text{x}\|^2} - 1\right), \ \epsilon > 0.
\end{align}
The smooth adjacency matrix containing the linkages between the MAPs, denoted by $\mathbf{A} = [a_{ij}] \in \mathbb{R}^{L \times L}$ can then be obtained as follows:
\begin{align}
a_{ij} =
\begin{cases}
\alpha_{\{\gamma,1\}} \left(\frac{ \| q_i - q_j\|_{\sigma}}{\|r\|_{\sigma}} \right), & \text{if } i \neq j,\\
0, & \text{if } i = j.
\end{cases}
\end{align}
Moreover, we assume that adjacent MAPs use different frequency channels to communicate and hence, do not interfere with each other. Therefore, the reliability of the communication links between the MAPs is directly reflected by the adjacency matrix $\mathbf{A}$. The degree matrix of the MAPs is defined by $\mathbf{D} = [d_{ij}] \in \mathbb{R}^{L \times L}$, where $d_{ij} = \mathbbm{1}_{i = j} \sum_{j = 1}^{L} \mathbbm{1}_{ \{a_{ij} > 0\}}, \forall i,j \in \mathcal{L}$, where $\mathbbm{1}_{\{.\}}$ denotes the indicator function.

\begin{figure}
  \centering
  \includegraphics[width=2.5in]{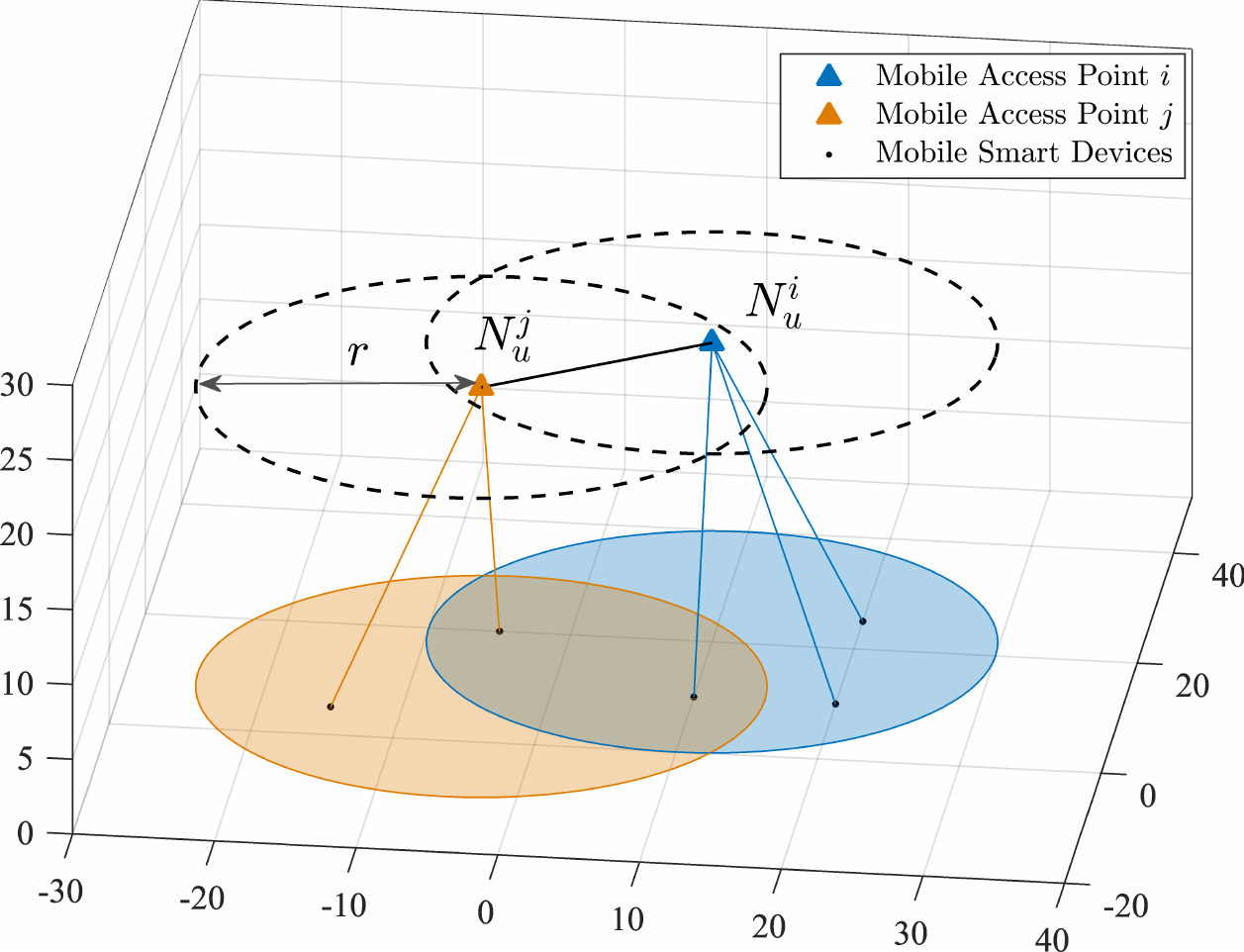}\\
  \caption{An example of two connected MAPs serving the underlying MSDs. The communication range of each MAP is depicted by the dotted lines while the area of influence is represented by the shaded circles.}\label{sys_model}
\end{figure}

The connectivity of MSDs is determined by their coverage by one of the available MAPs. A MAP has a certain area of influence under which it can reliably provide coverage to the MSDs. For simplicity, we assume the range of influence of the MAPs on the ground is the same as their communication range. Fig.~\ref{sys_model} illustrates a typical scenario of two adjacent MAPs providing connectivity to the MSDs inside their influence region enabling network-wide connectivity. An MSD $i$ is assumed to be connected to MAP $j$ if it is closer to it than any other MAP, i.e., $\|y_i - q_j\| < \|y_i - q_k\|, k \in \mathcal{L}\backslash j$, and the MAP has sufficient capacity to serve the MSD. The total number of MSDs connected to the MAPs are denoted by the vector $\mathbf{N}_u = [N_u^1, N_u^2, \ldots, N_u^L]^T$, while the maximum serving capacity of each MAP is denoted by $N^{\max}$. The distance based user association is motivated by the distance dependent signal decay. Each MSD aspires to connect to its nearest in-rage MAP unless constrained by the capacity of the host.

\vspace{-0.0in}
\section{Methodology} \label{Sec:Methodology}

In this section, we describe the methodology used to develop the cognitive and resilient connectivity framework for remotely deployed IoT devices. We assume that the locations of the MSDs are constantly changing and is beyond the control of the MAPs. Our objective is to autonomously configure the MAPs in a distributed manner to provide coverage to the MSDs as well as keeping them connected to other MAPs. The cognitive connectivity resilience framework can be summarized by the cognition loop illustrated in Fig.~\ref{cognition}. The individual blocks of the cognitive framework are elaborated in the subsequent subsections.

\subsection{MAP-MSD Matching}
At each iteration of the cognitive connectivity framework, there is an association between the MAPs and the MSDs. Since the wireless channel experiences distance dependent path loss, it is reasonable to assume a utility based on the Euclidean distance when an MSD gets served by an MAP. The utility of the matching between an MSD $i$ located at $y_i$ and an MAP $j$ located at $q_j$ is expressed as follows:
\begin{align}
\Phi(i,j) = - \| y_i - q_j \|^2,
\end{align}
The preference of MSD $i$ can be described as follows:
\begin{align}
v(i) = \underset{j \in \mathcal{L} : \|y_i - q_j\| < r}{\max} \{\Phi(i,j)\}.
\end{align}
Notice, that only the MSDs that are under the influence of the MAPs are matched and the uncovered MSDs remain un-matched. The optimal assignment results in the matrix $\boldsymbol{\varepsilon}$, where $\varepsilon_{ij} = \mathbbm{1}_{\{j = v(i)\}}, \forall i \in \mathcal{M}, j \in \mathcal{L}$. As a result, the number of MSDs matched to each MAP can be evaluated as $N_u^i = \sum_{j = 1}^{M} \varepsilon_{ij}, \forall i \in \mathcal{L}$.

\subsection{MAP Dynamics and Objective}
\begin{figure}
  \centering
  \includegraphics[width=3.4in]{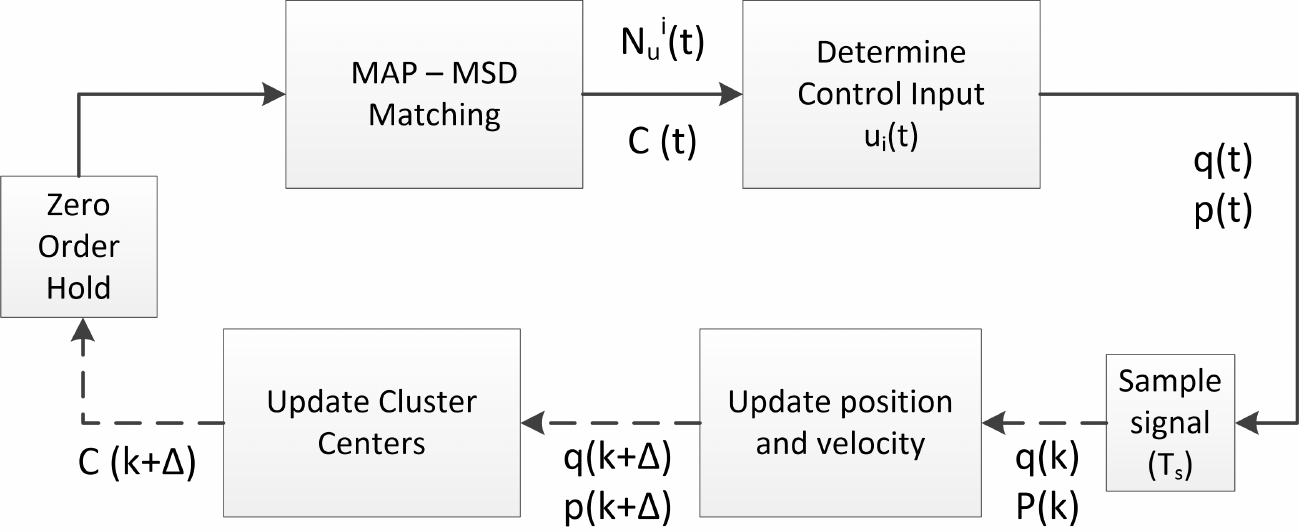}\\
  \caption{Feedback loop for cognitive connectivity resilience.}\label{cognition}
\end{figure}

We employ the kinematic model for the MAPs, whose dynamics can be written as follows:
\begin{align} \label{dynamics}
\dot{q}_i = p_i,\\ \notag
\dot{p}_i = u_i,
\end{align}
where $q_i, p_i, u_i \in \mathbb{R}^2, i \in \mathcal{L}$, represent the displacement, velocity, and acceleration of the devices respectively. For practical implementation, the displacement and velocity vectors are discretized with a sampling interval of $T_s$ before being updated with a step size $\Delta$ as shown in Fig.~\ref{cognition}. The goal is to design a control input $u_i$ for each MAP which eventually leads to a desired configuration. To achieve this, we build upon the framework developed in~\cite{flocking} for distributed multi-agent systems and provide modifications which leads to the desired behaviour of MAPs in the context of D2D wireless networks constrained by the underlying MSDs. To enhance spatial coverage, the MAPs should have less coverage overlap and should be spread out while remaining connected to other MAPs. Therefore, we define a minimum distance $0 \leq d < r$ such that two MAPs should not be closer to each other than $d$ unless they are forced to be closely located to serve a higher density of underlying MSDs.

%\subsection{Cognitive Connectivity Algorithm}
\subsection{Controller Design}

We propose a control input $\mathbf{u} = [u_1, u_2, \ldots, u_{L}]^T$ for each of the MAPs in the following form:
\begin{align}
u_i = f_i(\mathbf{q},\mathbf{A},\mathbf{N}_u) + g_i(\mathbf{p},\mathbf{A}) + h_i(\mathbf{q}, \mathbf{p}), \label{control_input}
\end{align}
where $f_i(\mathbf{q},\mathbf{A},\mathbf{N}_u)$ defines the gradient based term based on the attractive and repulsive forces between the MAPs, $g_i(\mathbf{p},\mathbf{A})$ is the velocity matching term that forces neighbouring MAPs to move with the same speed, and $h_i(\mathbf{q}, \mathbf{p})$ is the term defining the individual goals of each of the MAPs. Note that $\mathbf{q}$, $\mathbf{p}$, and $\mathbf{A}$ are functions of time. Each of the terms in~\eqref{control_input} are elaborated as follows:

\subsubsection{Attractive and Repulsive Function}
As highlighted earlier, the MAPs tend to maintain a minimum distance $d$ with other MAPs unless the serving capacity is exceeded. Therefore, a repulsive force is required from MAP $j$ to MAP $i$ if the distance between them is less than $d$ and an attractive force is needed from MAP $i$ to MAP $j$ if MAP $j$ exceeds the capacity of serving MSDs. In effect, MAP $i$ tends to share the load of MAP $j$ if it exceeds capacity in order to provide coverage to all the MSDs. Therefore, the $i^{th}$ element of $f_i(\mathbf{q},\mathbf{A},\mathbf{N}_u)$ can be defined as follows:
\begin{align}
f_i&(\mathbf{q},\mathbf{A},\mathbf{N}_u) = \sum_{j \in \mathcal{N}_i} \Bigg[ \Psi( \| q_j  - q_i \|_{\sigma} )  \ +  \notag  \\ & a \left( 1 - \alpha_{\{0,1\}} \left( \frac{\| (N_u^j - N^{\max})^+\|_{\sigma}}{\| N^{\max}\|_{\sigma}} \right) \right)  \Bigg] \mathbf{v}_{ij},i \in \mathcal{L},\label{grad_term}
\end{align}
where $\mathbf{v}_{ij} = \nabla \| q_j - q_i \|_{\sigma}$ represents the vector in the direction going from the MAP at location $q_i$ to the MAP at location $q_j$. The function $\Psi(z)$ is provided as follows:
\begin{align}
\Psi(z) = \alpha_{\{\gamma,1\}} \left( \frac{z}{\|r\|_{\sigma}} \right) \phi(z - \|d\|_{\sigma}),
\end{align}
where $\phi(z) = \frac{1}{2}[(a+b) \frac{(z+c)}{\sqrt{1 + (z+c)^2}} + (a - b)]$ is an un-even sigmoid function with $c = |a-b|/\sqrt{4ab}$ such that $\phi(z) \in (-a,a)$ and $\phi(0) = 0$.
Notice that the function $\Psi(z)$ is a product of two functions that results in the property that $\Psi(z) \leq 0$ if $z < \|d\|_{\sigma}$ and $\Psi(z) = 0$ otherwise, i.e., it provides a repelling force if MAP $i$ and MAP $j$ are closer than $d$ and is neutral if they are farther than $d$. Therefore, the first term in the multiplier of the gradient in \eqref{grad_term} ensures that the distance between neighbouring MAPs is at least $d$. The second term is related to the attraction between MAPs if the MSDs aspiring to connect to them are beyond their capacity, i.e., $N^{\max}$. The force depends on the number of unserved users $(N_u^j - N^{\max})^+$ normalized by the maximum capacity and is accomplished using the function $a(1 - \alpha_{\{0,1\}}(.))$ $\in [0,a)$, which is nonzero for strictly positive arguments.

\subsubsection{Velocity Consensus Function}
The velocity consensus function enables a matching between the velocities of neighbouring MAPs and is expressed as follows:
\begin{align}
g_i(\mathbf{p}, \mathbf{A}) = \sum_{j \in \mathcal{N}_i} a_{ij}(p_j - p_i), i \in \mathcal{L}.
\end{align}
The function implies that MAP $i$ tends to align its velocity to its neighbours weighted by the strength of the links.

\subsubsection{Individual Goal Function}
The individual goal function $h(\mathbf{q}, \mathbf{p})$ is defined as follows:
\begin{align}
h_i(\mathbf{q}, \mathbf{p}) = c_1 (q_i^r - q_i ) +  c_2 (p_i^r - p_i), i \in \mathcal{L},
\end{align}
where $q_i^r$ and $p_i^r$ are the reference position and velocity of MAP $i$, and $c_1$ and $c_2$ are positive constants denoting the relative aggressiveness to achieve the goal. If the goal of each MAP is precisely determined, the network can be appropriately configured. Assuming that each MAP is greedy to serve MSDs, a natural goal is to reach the centroid of the MSDs to allow a maximum number of MSDs to connect to it. Since the MSDs may be arbitrarily clustered, it is more efficient for the MAPs to move toward the centroid that is nearest to them. Hence the individual reference signals are selected as follows:
\begin{align}
q_i^r  =  C_i^*, \ \forall i \in \mathcal{L}, \notag\\
p_i^r  =  0, \ \forall i \in \mathcal{L},
\end{align}
where $C_i^*$ denotes the coordinates of the cluster center nearest to MSD $i$ and is further elaborated in the sequel. A reference velocity of $0$ implies that each MAP wants to eventually come to rest. The proposed cognitive loop propagates as follows: Given the spatial locations of the MSDs, a matching is made between the MAPs and the MSDs based on the distances and the maximum capacity of the MAPs. Based on the MAP-MSD association and the cluster centers (determined by the spatial location of MSDs), a control input is computed by each MAP independently and the system states of the are updated according to the dynamics provided by \eqref{dynamics}. After the locations and velocities are updated, new cluster centers of the MSDs are computed as their spatial locations might have changed due to mobility. Upon convergence, the control input becomes nearly zero and there is no further change in the configuration provided the MSDs do not change their positions.

\subsection{Cluster Centers}
In order to determine the destination of each MAP, we need information about the locations of the MSDs. Since the MSDs can move arbitrarily, they may not have a definitive spatial distribution. Therefore, it is reasonable to cluster the MSDs and use their centers as a potential destination for nearby MAPs. Assuming that the MSDs are partitioned into $K$ sets denoted by $\mathcal{S} = \{S_1, S_2, \ldots, S_K\}$, then the objective is to find the following:
\begin{align}
\underset{\mathcal{S}}{\arg \min} \sum_{i=1}^{K} \sum_{y \in S_i} \| y - \bar{C}_i\|^2,
\end{align}
where $\bar{C}_i$ denotes the centroid of the $i^{\text{th}}$ cluster. The optimal cluster centers are denoted by $\mathcal{C} = [C_1, C_2, \ldots, C_K]^T, C_i \in \mathbb{R}^2, \forall i = 1, \ldots, K$.
The solution to this problem can be obtained efficiently using Lloyd's algorithm~\cite{lloyds}. In our proposed model, the only centralized information needed is the coordinates of the cluster centers of the MSDs. It can either be obtained using a centralized entity such as the satellite or it can be locally estimated based on individual observations. However, if local observations are used, then a distributed consensus needs to be made regarding the final goal state of each MAP.
%It remains to be investigated if locally estimated cluster centers by each MAP results in a desirable network formation and is intended as a future work.

\vspace{-0.0in}
\section{Performance Evaluation}\label{Sec:Perf_evaulation}
In order to evaluate the performance of the proposed cognitive connectivity framework, we use the following metrics to measure the connectivity of the network:
\begin{enumerate}
\item \textbf{Proportion of MSDs covered:} It measures the percentage of total MSDs that are associated and served by one of the MAPs.
\item \textbf{Probability of information penetration in MAPs:} Assuming that the connected MSDs can communicate perfectly with the MAPs, the overall performance of the system depends on the effectiveness of communication between the MAPs using D2D links. One way to study the dynamic information propagation and penetration in D2D networks is based on mathematical epidemiology (See~\cite{iobt_junaid}). Since the network in this paper is finite, we make use of the $N$-intertwined mean field epidemic model~\cite{N_intertwined} to characterize the spread of information. The steady state probability of MAP $i$ being informed by a message, denoted by $\nu_{i \infty}$, propagated in the D2D network at an effective spreading rate of $\tau$ is bounded as follows (See Theorem 1 in~\cite{N_intertwined}):
    \begin{align}
    0 \leq \nu_{i \infty} \leq  1 - \frac{1}{1 + \tau d_{ii}}.
    \end{align}
\item \textbf{Reachability of MAPs:} While it is important to determine the spreading of information over the D2D network, it is also crucial to know whether the D2D network is connected or not. It can be effectively determined using the algebraic connectivity measure from graph theory, also referred to as the \emph{Fiedler value}, i.e., $\lambda_2(\mathbf{L})$, where $\lambda_2(.)$ denotes the second-smallest eigenvalue and $\mathbf{L}$ is the Laplacian matrix of the graph defined by the adjacency matrix $\mathbf{A}$. A nonzero Fiedler value indicates that each MAP in the network is reachable from any of the other MAPs. The Laplacian matrix is defined as follows:
\begin{align}
\mathbf{L} = \mathbf{D} - \mathbf{A}.
\end{align}
\end{enumerate}
These metrics complement each other in understanding the connectivity of the network. The resilience, on the other hand, is measured in terms of the percentage of performance recovery after an event of failure has occurred.
%Information spread in contact based networks can be modeled by the epidemic models used for studying the spread of viruses.
%How effectively can information be propagated into the network

\begin{table}[]
\centering
\caption{Simulation Parameters.}
\label{parameters}
\begin{tabular}{|l|l|l|l|}
\hline
Parameter & Value & Parameter & Value \\ \hline
$M$         & 2000  & $a$         & 5     \\ \hline
$L$         & 80    & $b$         & 5     \\ \hline
$r$         & 24    & $c_1$        & 0.2   \\ \hline
$d$         & 20    & $c_2$        & 0.1   \\ \hline
$\epsilon$         & 0.1   & $s$         & 0.2   \\ \hline
$N^{\max}$      & 80    &    $\tau$       &  1     \\ \hline
$h$      & 20    &       $T_s$    &   0.01   \\ \hline
$k$      & 3    &    $\gamma$      &    0.2  \\ \hline
\end{tabular}
\end{table}

\begin{figure*}[t!]
\centering
\subfloat[]{\includegraphics[width = 2.4in]{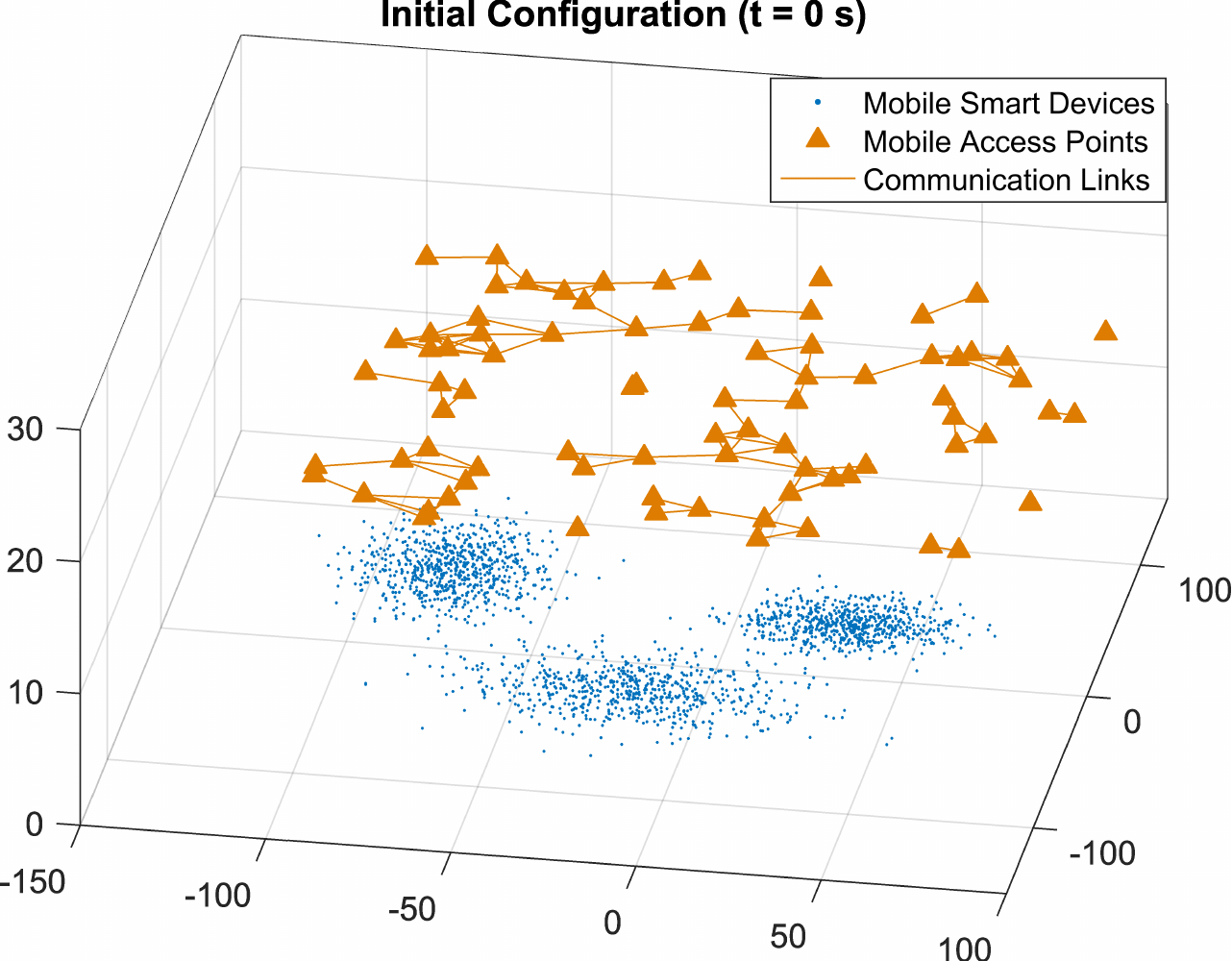} \label{initial_config}}
\subfloat[]{\includegraphics[width = 2.4in]{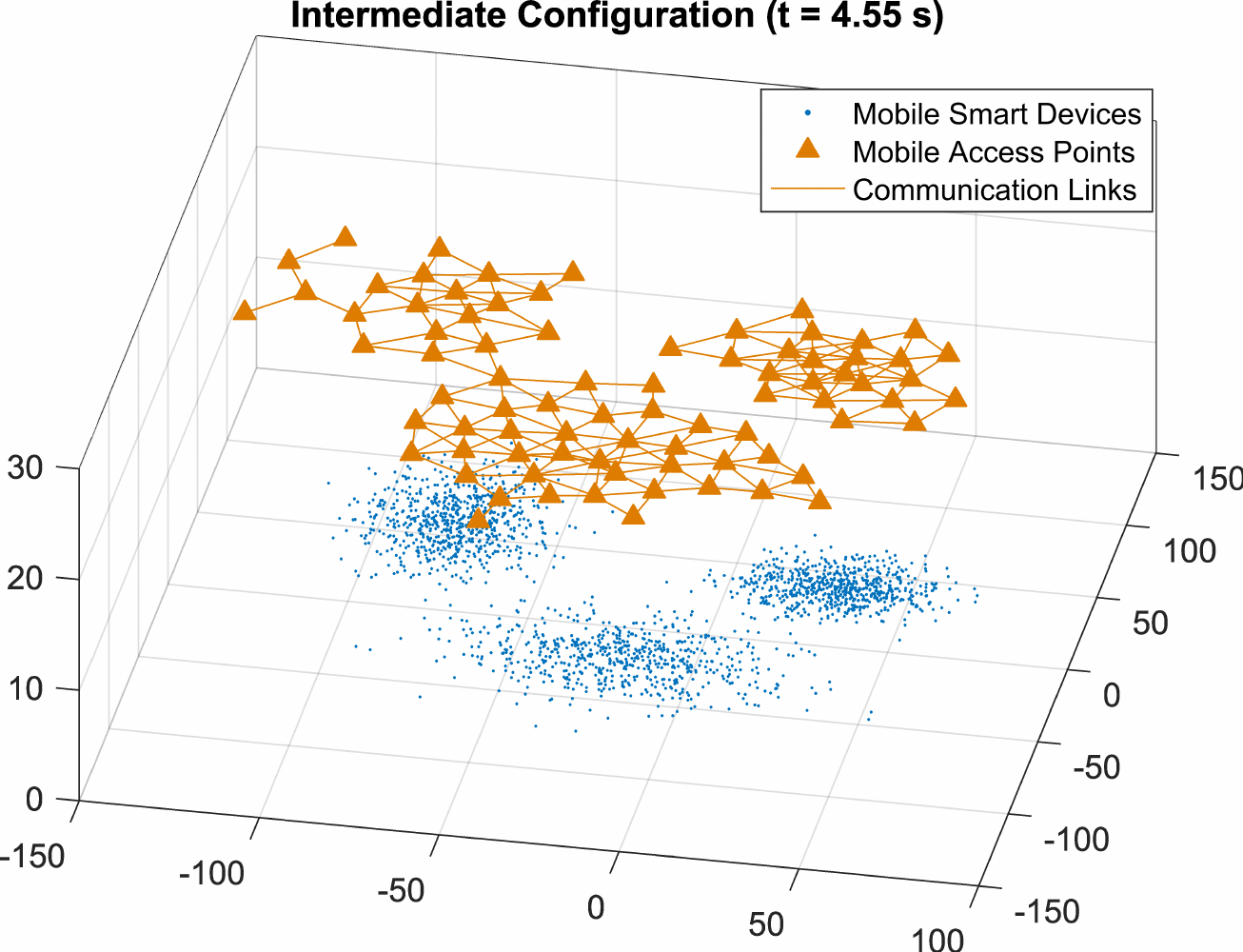} \label{interim_config}}
\subfloat[]{\includegraphics[width = 2.4in]{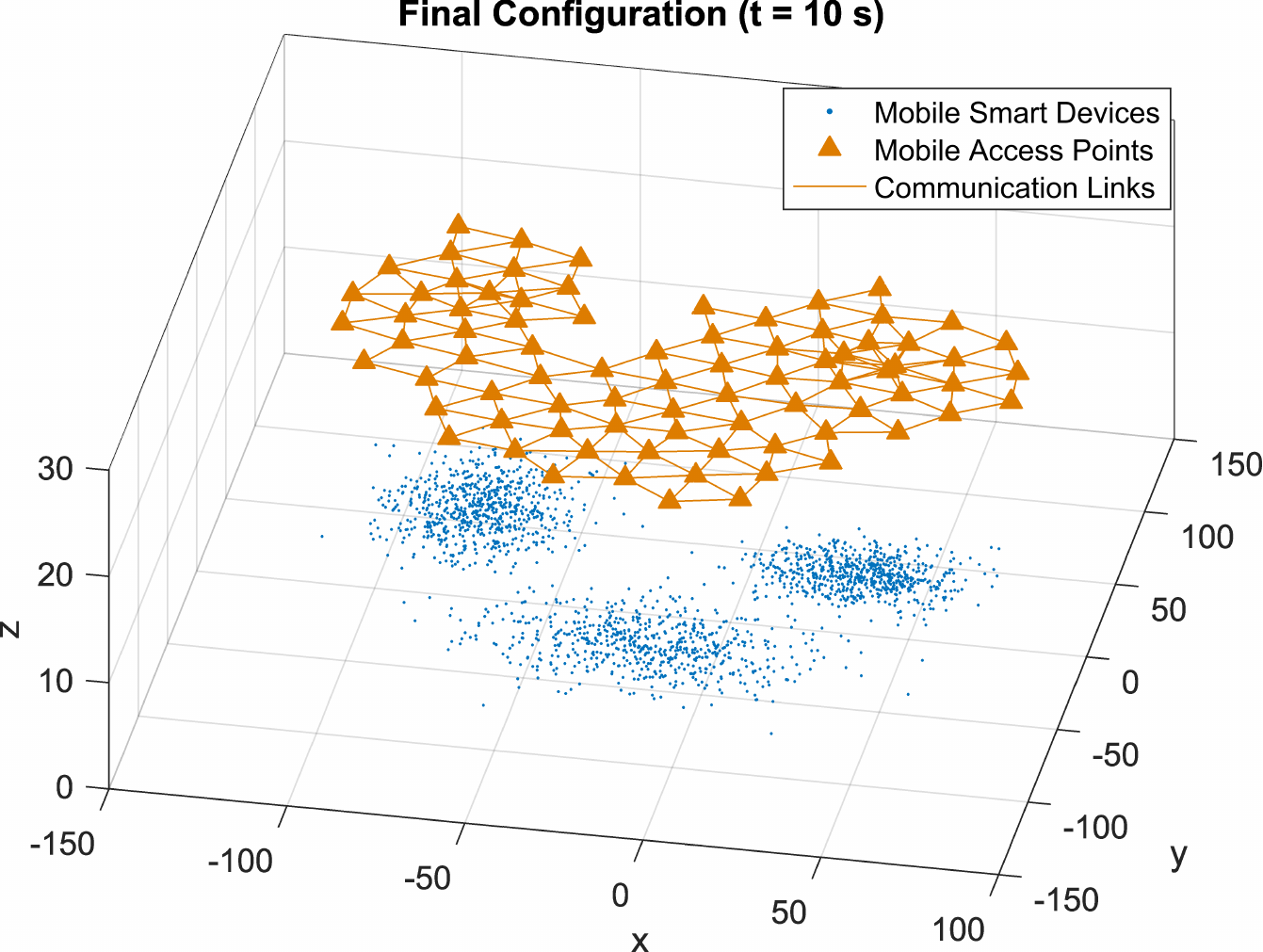} \label{final_config}}\\
\subfloat[]{\includegraphics[width = 2.4in]{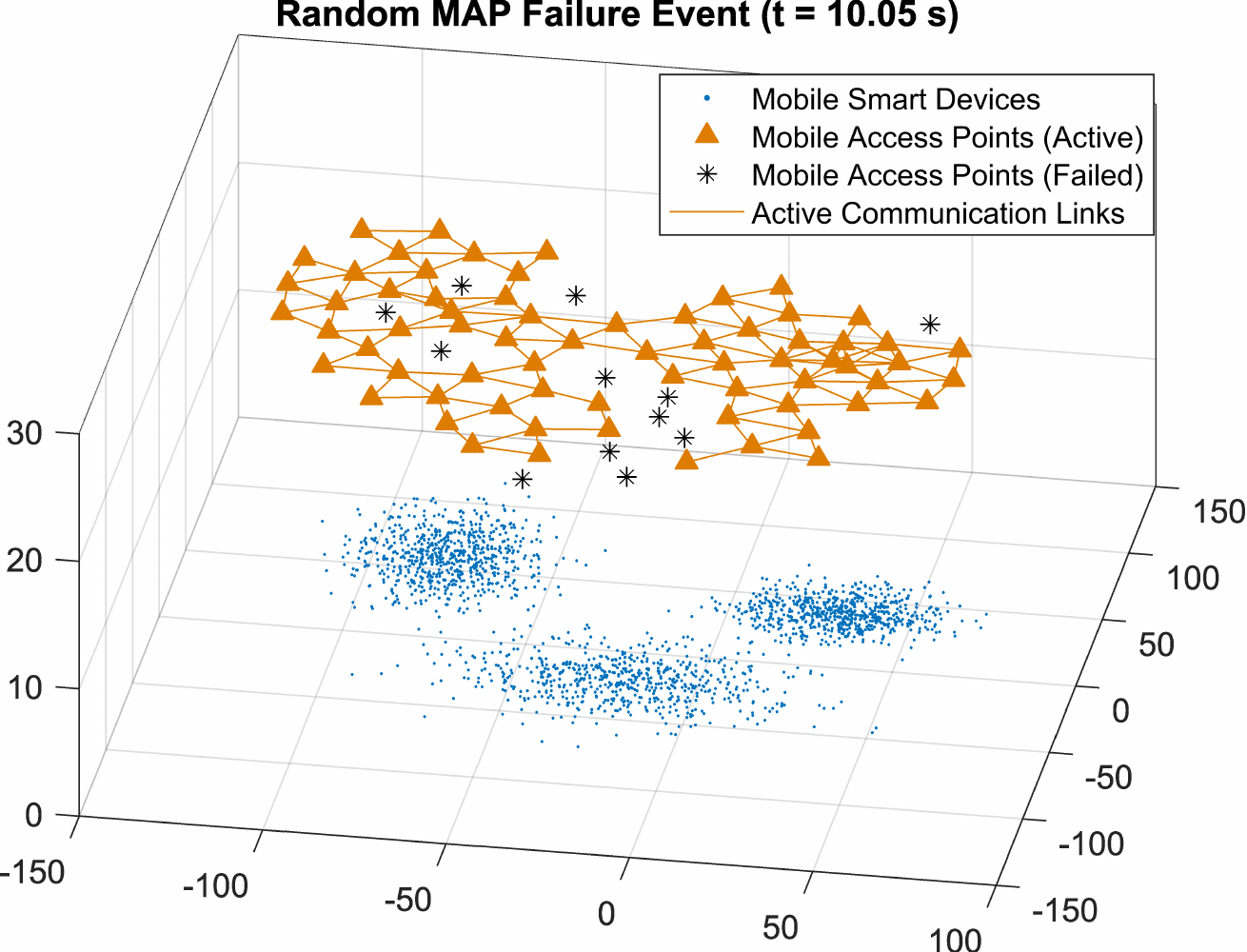} \label{failure_config}}
\subfloat[]{\includegraphics[width = 2.4in]{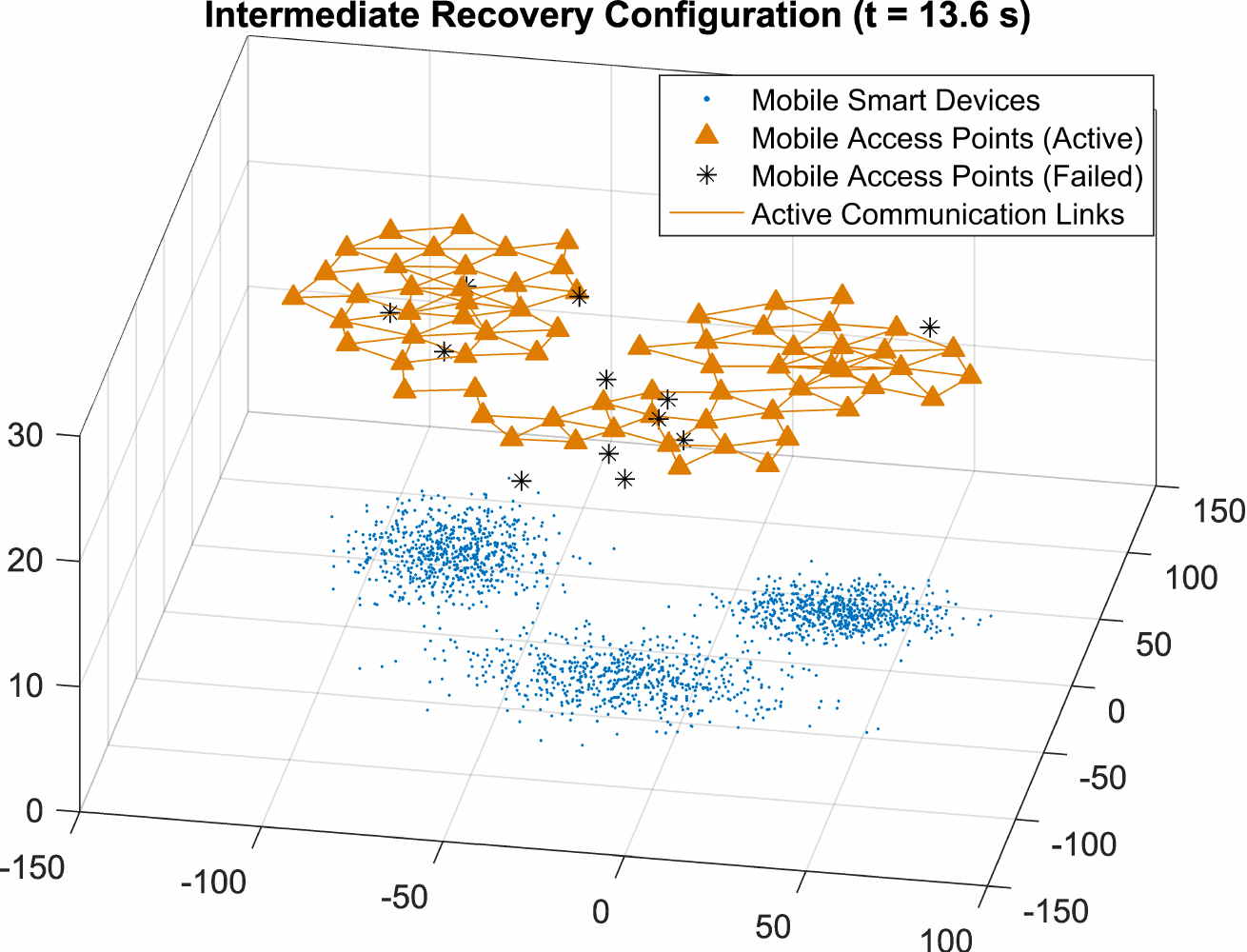} \label{interim_recovery}}
\subfloat[]{\includegraphics[width = 2.4in]{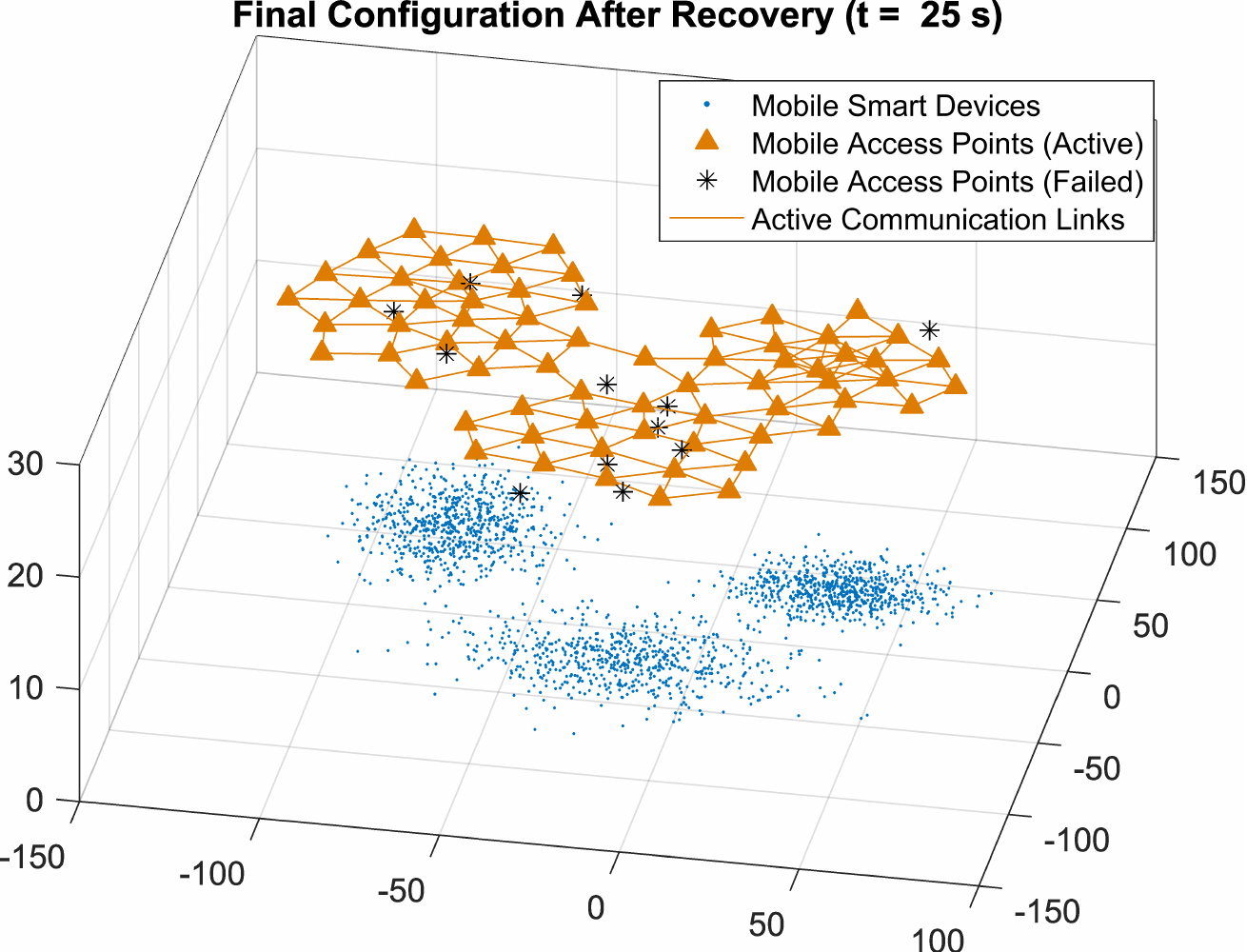} \label{recovery_config}}
\caption{Example run of the cognitive algorithm showing the initial intermediate and final configuration after convergence. A random MAP failure event occurs at $t = 12$ s, making $20$\% of the MAPs unavailable. The cognitive framework adaptively re-configures itself to improve network connectivity.}
\label{results}
\end{figure*}

\vspace{-0.0in}
\section{Results \& Discussion} \label{Sec:Results}

In this section, we first describe the simulation settings before providing results on the performance of our proposed cognitive connectivity framework. We assume a bi-layer communication network comprising of MSDs such as IoT devices and MAPs such as UAVs. The $M$ MSDs are distributed in $\mathbb{R}^2$ according to a 2-D Gaussian mixture model with equal mixing proportions. The mean vectors are $\mu_1 = [30,40]^T, \mu_2 = [-20,-20]^T$, and $\mu_3 = [-80,60]^T$, and the covariance matrices are as follows:
\begin{align}
\hspace{-0.3cm}\Sigma_1 = \hspace{-0.1cm}
\arraycolsep=1.5pt\def\arraystretch{1.4}
  \left[ {\begin{array}{cc}
   200 & 0 \\
   0 & 100 \\
  \end{array} } \right],
\Sigma_2 = \hspace{-0.1cm}
  \left[ {\begin{array}{cc}
   500 & 0 \\
   0 & 200 \\
  \end{array} } \right],
  \Sigma_3 = \hspace{-0.1cm}
  \left[ {\begin{array}{cc}
   150 & 0 \\
   0 & 300 \\
  \end{array} } \right]\hspace{-0.1cm}.
\end{align}
The mobility of the MSDs is modeled by a scaled uniform random noise at each time step, i.e., $y_i(t+1) = y_i(t) + s\xi$, where $\xi \sim$ Uniform($[-1,1] \times [-1,1]$), where $s$ represents the scale. The $L$ MAPs are initially distributed uniformly in the plane perpendicular to the vector $[0,0,h]^T$. The initial velocity vectors of the MAPs are selected uniformly at random from the box $[-1,-2]^2$. A list of all the remaining parameter values used during the simulations is provided in Table~\ref{parameters}. We run the simulations with a step size of $\Delta = T_s$ up to $t = 25$ s.

Fig.~\ref{initial_config} shows the initial configuration at $t = 0$, when the MAPs are randomly deployed over an underlying population of MSDs. The MSDs keep moving with time in random directions as observed by the different MSD locations in Fig.~\cref{initial_config,interim_config,final_config}. As the proposed cognitive connectivity framework evolves, the MAPs tend to move toward the closest group of MSDs as shown in Fig.~\ref{interim_config}. Finally, when the framework converges, the MAPs develop a desirable connected formation hovering over the MSDs as shown by Fig.~\ref{final_config}. It should be noted that the MAPs are located closer to each other in areas where MSDs are densely deployed, such as around cluster centers. In areas where the MSDs are sparsely located, the MAPs develop a regular formation.

Next, we investigate the impact of a random MAP failure event on the connectivity of the network and evaluate the response of the proposed cognitive framework in such a situation. Fig.~\ref{failure_config} shows an induced random failure of 20\% of the MAPs, which results in loss of coverage to the MSDs as well as reduction in the connectivity of the MAPs. The proposed framework immediately starts responding to the coverage gap created by the MAP failure and tends to reconfigure itself as shown by the intermediate snapshot in Fig.~\ref{interim_recovery}. Eventually, the framework converges leading to a coverage maximizing configuration while maintaining connectivity of the MAPs as shown in Fig.~\ref{recovery_config}.

%Fig. is useful as it shows that at higher failure proportion, the MAP network becomes disconnected.

\begin{figure}
  \centering
  \includegraphics[width=2.7in]{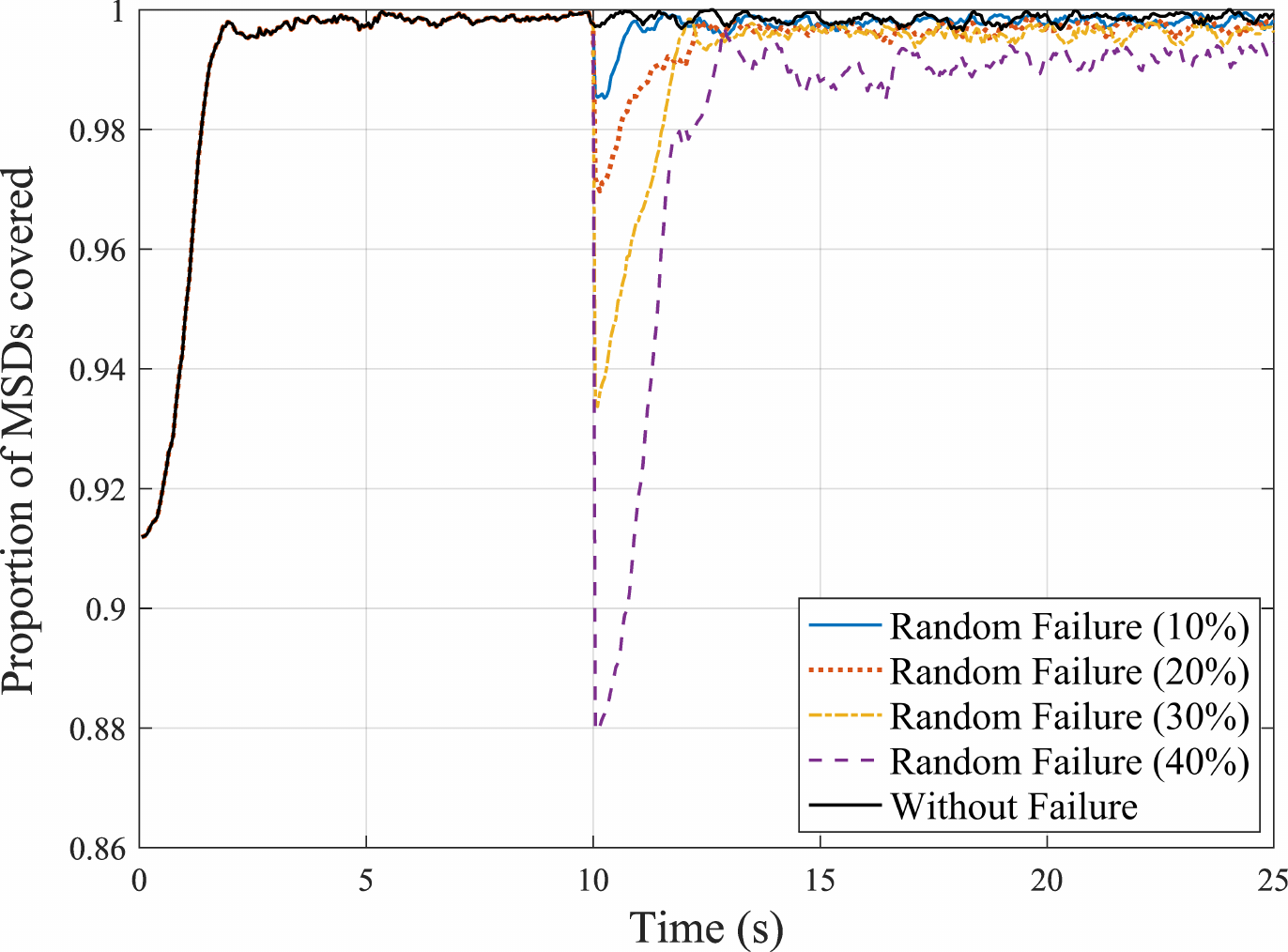}\\
  \caption{Proportion of MSDs covered by the MAPs.}\label{pc_covered}
\end{figure}

\begin{figure}
  \centering
  \includegraphics[width=2.7in]{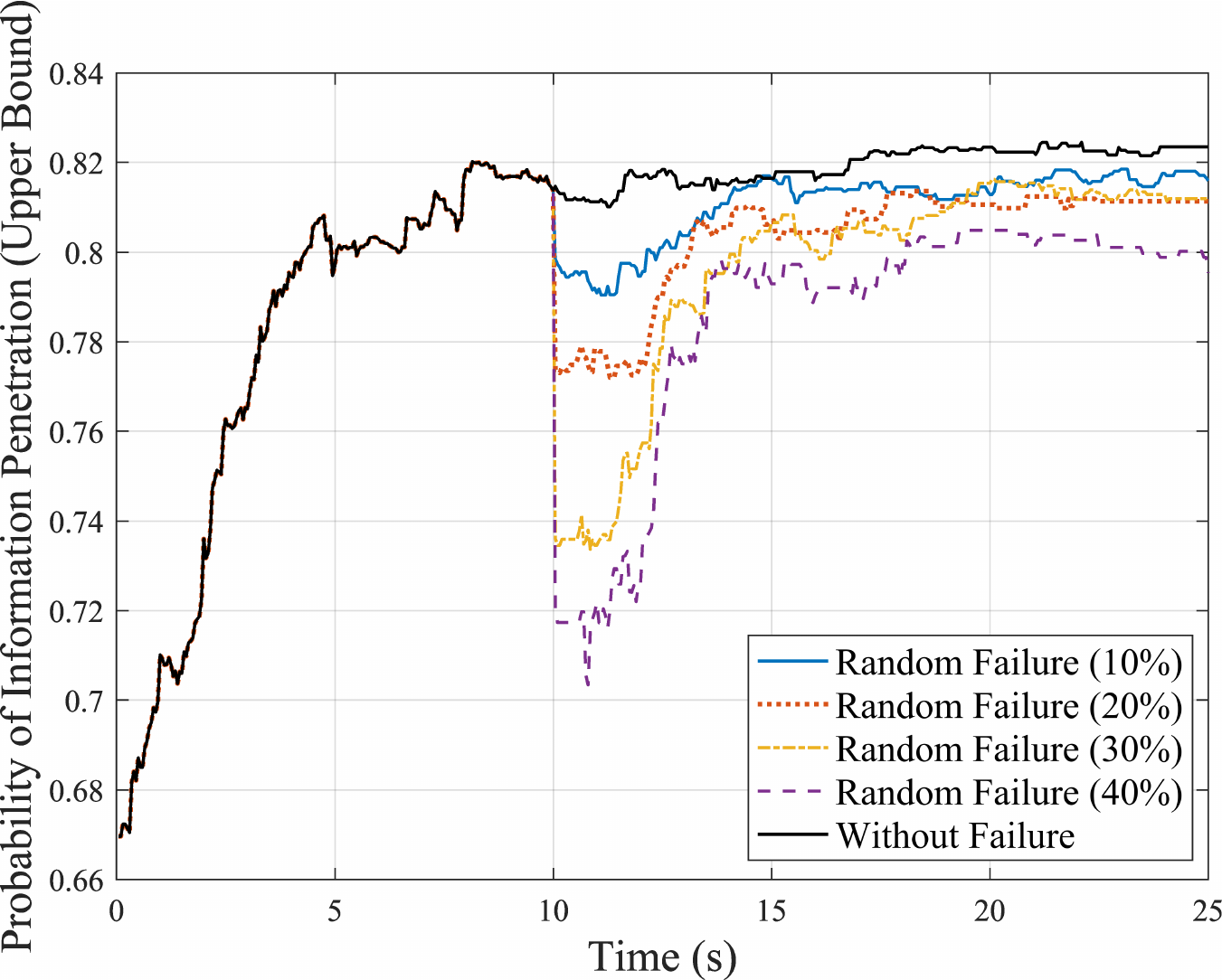}\\
  \caption{Probability of information penetration in the D2D enabled MAP network.}\label{pc_informed}
\end{figure}

\begin{figure}[t]
  \centering
  \includegraphics[width=2.7in]{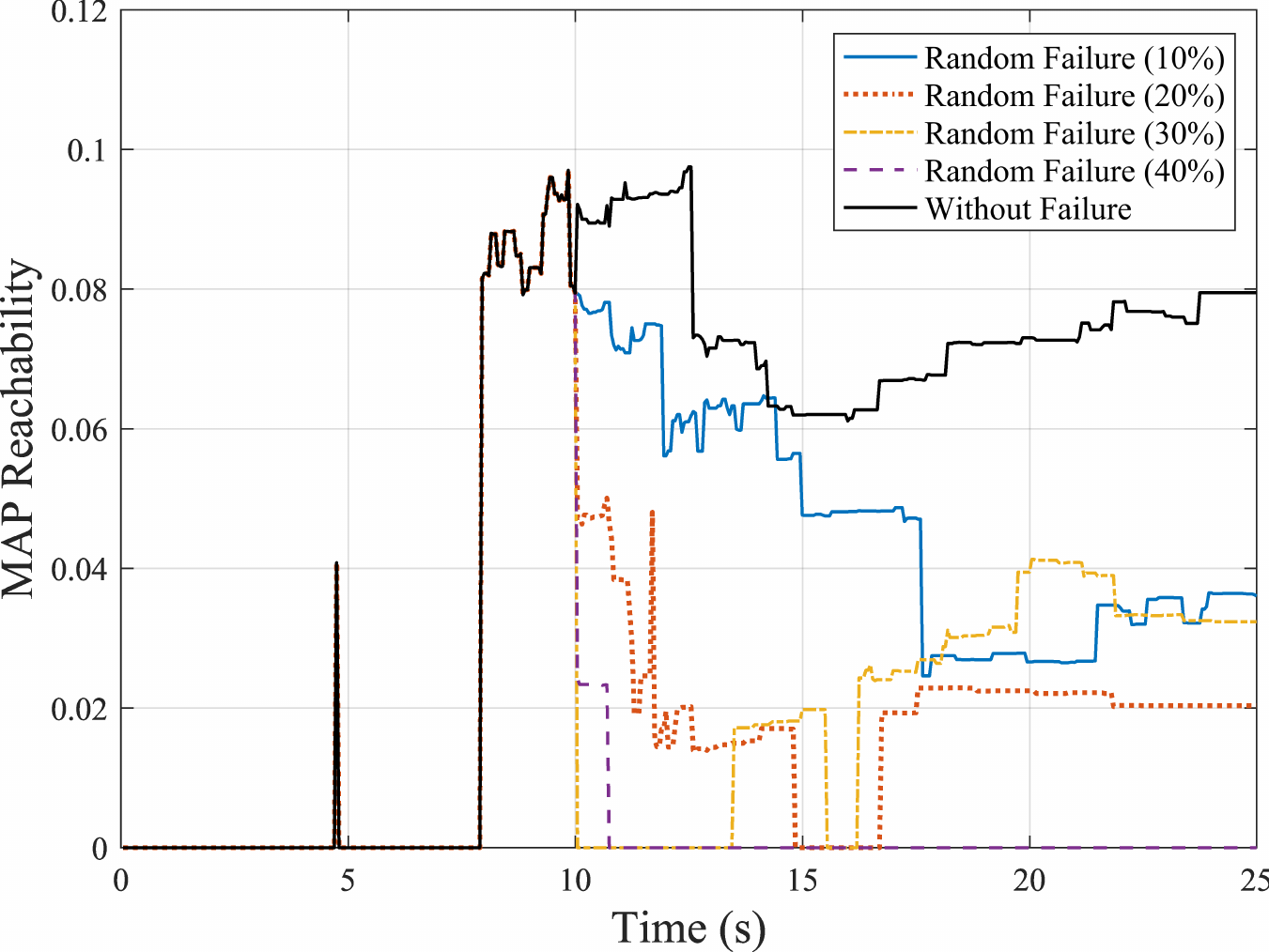}\\
  \caption{Reachability of MAPs determined by the algebraic connectivity.}\label{algebraic}
\end{figure}

Finally, we test the resilience of the proposed framework under different levels of MAP failure events. We simulate varying severity of device failure events from 10\% to 40\%  failed MAPs in the overlay network. Fig.~\ref{pc_covered} illustrates the proportion of MSDs that are covered by the MAPs. Without any failure, it is observed that the proposed framework successively improves the coverage until almost all the MSDs are covered. Notice that the coverage fluctuations occur due to the continuously mobile MSDs. Once the device failure event occurs at around $t = 10$ s, the framework responds and is able to quickly restore maximum coverage except in the case of 40\% failure, in which the coverage is not fully restored. In Fig.~\ref{pc_informed}, we plot the probability of information penetration in the MAP network. It can be observed that the framework is able to recover up to 97\% of the original value. However, it is important to note that the probability of information dissemination is an upper bound and does not provide information about the reachability of the MAPs. In this situation, Fig.~\ref{algebraic}, which shows the algebraic connectivity of the MAPs, proves to be extremely useful. It is observed that when the failure proportion is 10\%, 20\%, or 30\%, the reachability can be restored by the proposed framework, as indicated by the nonzero algebraic connectivity. However, in the event of 40\% failure, the algebraic connectivity remains zero even after reconfiguration, which implies that the MAP network is no more connected. However, since the probability of information penetration is still high, it implies that the MAPs have also clustered around the MSD clusters thus providing effective intra-cluster connectivity.

\vspace{-0.0in}
\section{Conclusion} \label{Sec:Conclusion}
In this paper, we present a cognitive connectivity framework that is able to reconfigure itself autonomically in a distributed manner to interconnect spatially dispersed smart devices thus enabling the Internet of things in remote environments.
Resilience of connectivity has been investigated in response to the mobility of the underlay network as well as random device failures in the overlay network.
It is shown that if sufficient number of overlay devices are available, then the developed distributed framework leads to high network connectivity which is resilient to mobility and device failures. However, if sufficient overlay devices are not deployed, the framework tends to provide connectivity locally to the devices in each cluster of the underlay network.

\vspace{-0.1in}
\bibliographystyle{IEEEtran}
\bibliography{new_references}

% that's all folks
\end{document}